\documentclass[10pt]{article}
\pdfoutput=1

\usepackage{pscc}
\usepackage{graphicx}

\begin{document}

%\vspace{-0.2cm}
%\begin{center} [16th Power Systems Computation Conference, Glasgow, Scotland, July 14-18, 2008 (PSCC'08)]
%\end{center}
\vspace{0.2cm}

\papertitle{Reliability Analysis of Electric Power Systems\\ Using an Object-oriented Hybrid Modeling Approach}

\noindent \names{\large Markus  Schl{\"a}pfer, Tom Kessler, Wolfgang Kr{\"o}ger}\endnames \names{Swiss Federal Institute of Technology}\endnames
\names{Zurich, Switzerland}\endnames \names{schlaepfer@mavt.ethz.ch}\endnames

\begin{multicols}{2}

\abstr{The ongoing evolution of the electric power systems brings about the need to cope with increasingly complex interactions of technical
components and relevant actors. In order to integrate a more comprehensive spectrum of different aspects into a probabilistic reliability assessment
and to include time-dependent effects, this paper proposes an object-oriented hybrid approach combining agent-based modeling techniques with
classical methods such as Monte Carlo simulation. Objects represent both technical components such as generators and transmission lines and
non-technical components such as grid operators. The approach allows the calculation of conventional reliability indices and the estimation of
blackout frequencies. Furthermore, the influence of the time needed to remove line overloads on the overall system reliability can be assessed. The
applicability of the approach is demonstrated by performing simulations on the IEEE Reliability Test System 1996 and on a model of the Swiss
high-voltage grid.}

\vspace{-0.4cm}

\section{Introduction}
%\vspace{-0.2cm}
\PARstart{T}{he} ongoing evolution of the electric power systems due to market liberalization and the integration of distributed generation is
leading to increasingly complex and hard-to-predict interactions of technical components, relevant actors and the operating environments.
Furthermore, recent large-area blackouts in North America and Europe demonstrated the potential consequences of inadequate operator response times to
contingencies (e.g. \cite{USBO:2004}). In recent years several advanced methods have been developed to assess the reliability of electric power
systems in general and to model and analyze cascading blackouts (e.g. \cite{Chen:2005,Rios:2002}). However, these approaches do not explicitly simulate the evolution of the events in time and represent the operator intervention to contingencies by using highly simplified models not taking into account the time needed for the corrective action. While Anghel et al. \cite{Anghel:2007} introduce a time-dependent probabilistic
approach incorporating a model for the utility response to line overloads, the influence of the response time on the occurrence of cascading line outages remains neglected.

The contribution of this paper is to present a basic modeling framework which allows the explicit integration of highly nonlinear, time-dependent effects and
non-technical factors into a probabilistic reliability assessment. Therefore, we apply an object-oriented hybrid approach combining agent-based
modeling techniques \cite{Inverno:2004} with classical methods such as Monte Carlo simulation \cite{Billinton:1994}. Objects represent both technical
components such as generators and transmission lines and non-technical components such as grid operators. They interact with each other directly
(e.g. via the generator dispatch) or via the physical power flows on the network. By means of long-term simulations the statistical data is gathered
for the calculation of system reliability indices and for the estimation of blackout frequencies.

The paper is organized as follows. Section 2 introduces the conceptual basics of the modeling framework and the derivation of the different
component models. In section 3 we present the results of applying the model to the IEEE Reliability Test System 1996 and to the Swiss high-voltage
system. Section 4 concludes.
%\vspace{-0.2cm}

\section{Modeling Framework}
%\vspace{-0.2cm}

\subsection{Conceptual Basics}
%\vspace{-0.2cm}

The conceptual modeling framework consists in the abstraction of the relevant technical and non-technical components of the electric power system as
individual interacting objects. Each object is modeled by attributes and rules of behavior. An example for an attribute is a technical component
constraint such as the rating of a transmission line. The rules of behavior are represented by using finite state machines (FSM) and include both
deterministic and stochastic time-dependent, discrete events. A deterministic event is, for instance, the outage of a component when reaching a
failure threshold, while stochastic processes are probabilistic component failure models using Monte Carlo techniques. The integration of
non-technical components is demonstrated by modeling the behavior of the grid operators in case of line overloads. For the corresponding interactions
between the operators and the technical components we make use of agent-based modeling techniques. Furthermore, we account for the possible division of the power system into several control areas. To each control area a distinct grid operator and a distinct control object are
assigned. The control object is not an abstraction of a technical component as such but rather represents an implementation construct which controls
the balance between generation and load within the corresponding control area. The model captures the system behavior over an operational period of one year.

\subsection{Component Models}

The components of the power system as being modeled as objects are $n_L$ loads, $n_G$ generators, $n_T$ transmission lines, $n_B$ busbars and $n_K$
grid operators.

\subsubsection{Loads}
%\vspace{-0.2cm}

The power demand trajectory $D_i(t)$ of load $i$ is described by:
\begin{equation}\label{eq:loadReconnection}
D_i(t) = \gamma(t)D_i^{max}(1+\rho(t)) - \Delta D_i(t)
\end{equation}
The demand factor $\gamma(t)$ is the actual time-dependent percentage of the peak demand $D_i^{max}$ and follows a chronologically changing load
profile over the predefined time period of one year. The percent deviation $\rho(t)$ represents stochastic demand fluctuations and is sampled hourly assuming a normal distribution with $\rho(t) \sim N(0,\sigma^2)$ and standard deviation $\sigma = 0.0192$ according to \cite{Dai:2001}. The value of $\rho(t)$ is assumed to be equal for all loads within the same control area. The term $\Delta D_i(t)$ represents the actual amount of partially shed load.

Figure \ref{fig:FSMloads} shows the FSM as implemented for the load objects. With the exception of the restoration process all the
transitions of the four-state model are externally governed by the control object $k$ of the corresponding control area.

\begin{figurehere}
\centering
\includegraphics[width=2.8in]{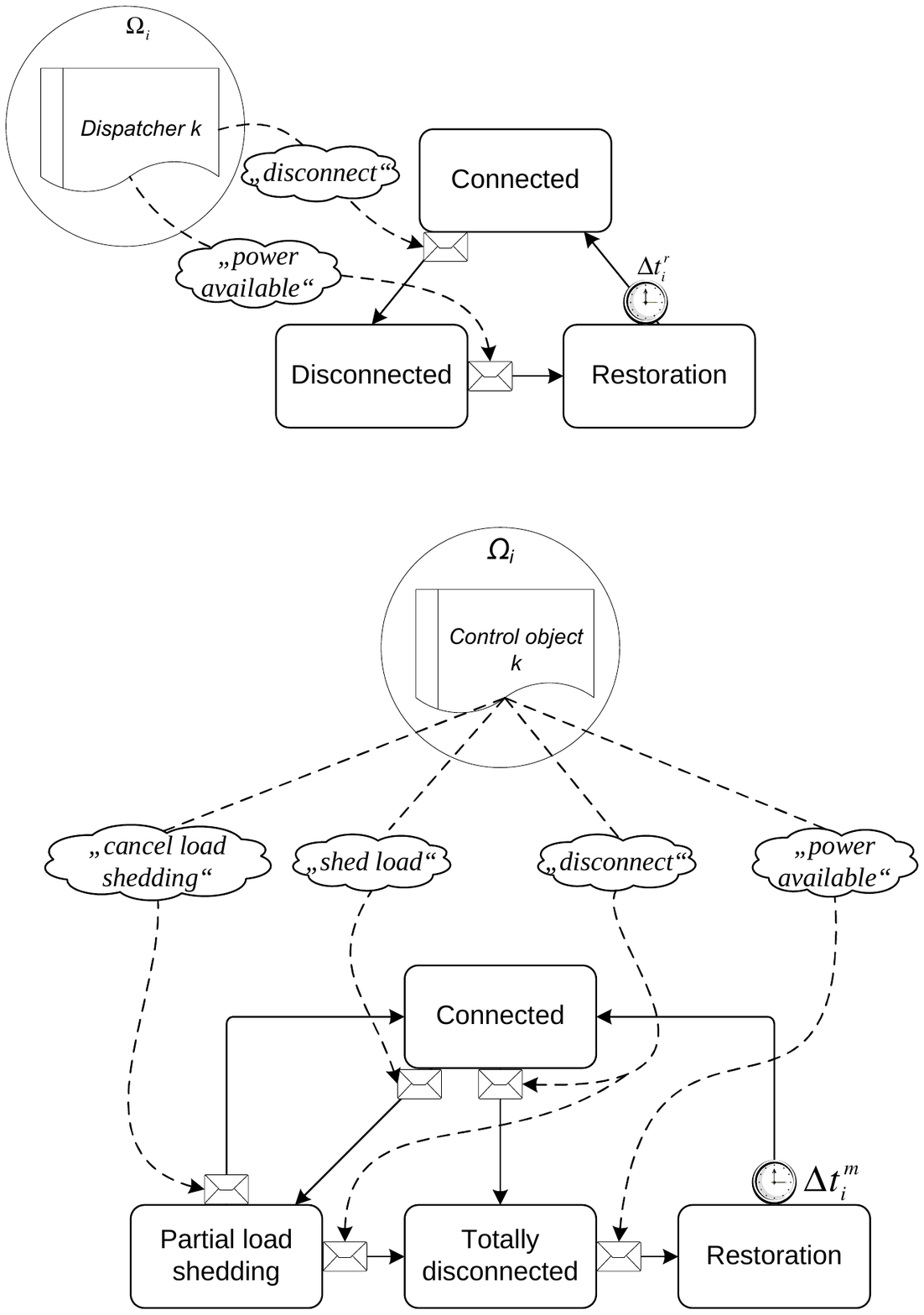}
\caption{Finite state machine for the load objects} \label{fig:FSMloads}
\end{figurehere}
\vspace{+0.2cm}

Partial load shedding occurs only when the control object sheds load due to an operator action for removing a line overload. As soon as the
transmission system can be operated within its security margins again, the load object receives the signal to cancel the partial load shedding. The
load gets totally disconnected if there is not enough generation capacity available within the entire system to cover its demand or as a consequence
of system splitting \mbox{(see section \ref{section:Splitting})}. If several loads have to be disconnected within one control area all loads are given the
same priority to be shed and are therefore selected randomly.

The restoration process is started once enough generation capacity is available again to cover the disconnected demand, and is modeled by a
queue technique. The load which has been disconnected first is also restored first, the subsequent one waits until the previous is reconnected. Based on \cite{Kirschen:2003} we assume an incremental overall restoration rate $\nu ( \Delta t^m_{tot})$ for four different restoration stages according to table \ref{tab:restRates}, where $\Delta t^m_{tot}$ is the elapsed time measured from the start of the overall restoration process $m$. Hence, the
time needed to reconnect a specific load, $\Delta t_i^m$, is dependent on the actual overall restoration stage.

\vspace{0.4cm}
\begin{tablehere}
\centering
\begin{tabular}{| l | l |}\hline\hline
$\Delta t^m_{tot}$ [min] & $\nu(t)$ [MW/min] \\\hline 0-30 & 10.0 \\ 30-60 & 33.3 \\ 60-90 & 66.6 \\ $>$ 90 & 83.3 \\\hline\hline
\end{tabular}
\vspace{0.2cm}
\caption{Stages of the overall restoration process and corresponding load restoration rates, adopted from \cite{Kirschen:2003}} \label{tab:restRates}
\end{tablehere}
\vspace{0.4cm}

\subsubsection{Generators}
%\vspace{-0.2cm}

The commitment of the generating units is continuously governed by the control object in order to cover the actual demand $D_{k}(t) =
\sum_{i\in \Omega_k}D_i(t)$ within the respective control area $k$. Being constrained by the maximum power outputs $P_j^{max}$, the commitment
and economic dispatch follows a heuristic priority list method according to \cite{Wood:1996} and is implemented in the control object. By using a recursive algorithm and starting with the highest
priority, $D_k(t)$ is equally distributed among the units with the same priority. As their maximum capacity is reached the algorithm proceeds
to the next lower priority and so forth. In case $D_k(t)$ is larger than the available generation capacity within a control area, the control
object analogously commits available generating units from the other control areas of the system. As a simplification, ramp rates and maintenance
are not considered at this stage of our work. The FSM for the generator object is made up of a two-state model for repairable forced failures being treated as random events as shown in figure \ref{fig:FSMgenerators}.

\begin{figurehere}
\centering
\includegraphics[width=2.0in]{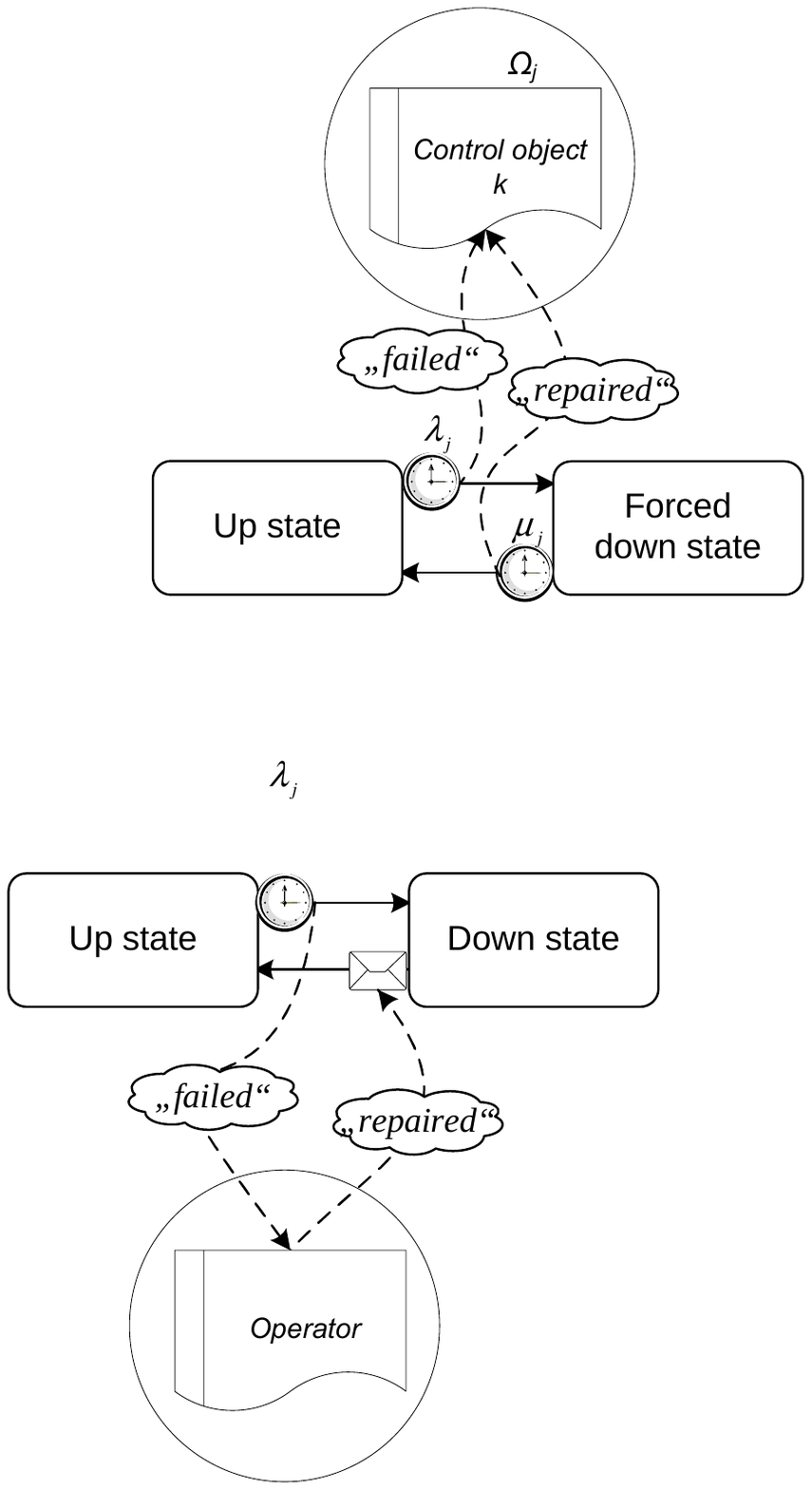}
\caption{Finite state machine for the generator objects} \label{fig:FSMgenerators}
\end{figurehere}
\vspace{+0.2cm}

The repairable forced failures are modeled by an independent stochastic up-down-up cycle assuming stationarity and constant failure and repair rates
$\lambda_j=1/MTTF_j$ and $\mu_j=1/MTTR_j$ respectively. Hence, this alternating renewable process is characterized by the cumulative distribution
functions of the failure-free times $\tau_j$ and repair times $\tau_j^{'}$ respectively, and by the probability $p_j$ that the generating unit is in
upstate at $t=0$ \cite{Birolini:2007}:
\begin{equation}\label{eq:genF}
F_j(t_u) = Pr\{\tau_j\leq t_u\} = 1-e^{-\lambda_j t_u}
\end{equation}
%\vspace{-1.2cm}
\begin{equation}\label{eq:genG}
G_j(t_u) = Pr\{\tau_j^{'}\leq t_d\} = 1-e^{-\mu_j t_d}
\end{equation}
%\vspace{-1.2cm}
\begin{equation}\label{eq:genp}
p_j = \frac{\mu_j}{\lambda_j+\mu_j}
\end{equation}
%\vspace{-1.2cm}

where $t_u$ and $t_d$ are the time spans measured from the moment of entering the upstate and forced down state respectively. All state transitions
are reported to the control object of the corresponding control area.

\subsubsection{Transmission lines}
%\vspace{-0.2cm}

The time variant line flows are calculated by the DC power flow approximation with \mbox{$P_\ell (t) =
x_{ab}^{-1}\big(\theta_{a}(t)-\theta_{b}(t)\big)$}, where $P_\ell(t)$ is the active power flow on line $\ell$ connecting busbar $a$ with busbar $b$, having reactance $x_{ab}$ and phase angles $\theta_{a}(t)$ and $\theta_{b}(t)$. The approximate solution of the power flow problem does not allow to analyze voltage disturbances. Nevertheless, we assume the DC model to be appropriate for analyzing cascading events due to line overloads and for showing the feasibility of the proposed modeling concept.

A five-state model for the basic behavior of the transmission line is used considering outages triggered by its protection device and by independent random failures, see figure \ref{fig:line}. Thereby, the protection device is modeled by a separate FSM.

\begin{figurehere}
\centering
\includegraphics[width=2.6 in]{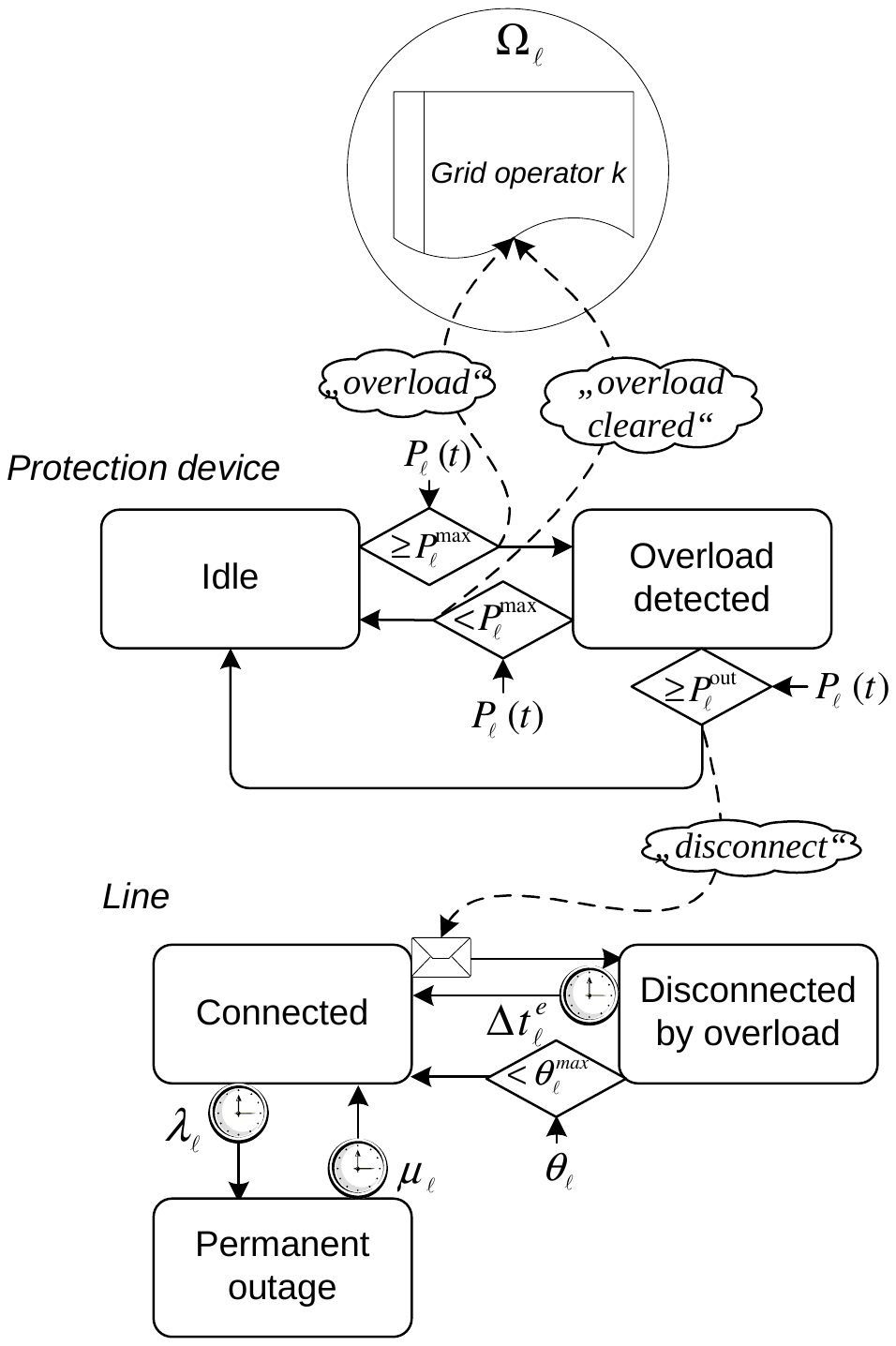}
\caption{Finite state machine for the transmission line objects} \label{fig:line}
\end{figurehere}
\vspace{+0.2cm}

In our model the protection device has two functions. Firstly, it continuously measures the power flow $P_\ell(t)$ and sends an alarm message ("overload") to the operator of the control area if $P_\ell(t)$ becomes equal to or larger than the line rating $P_\ell^{max}$. Secondly, if $P_\ell(t)$ reaches $P_\ell^{out}$ it disconnects the line. However, as a consequence of the stochastic time-dependent system behavior or the intervention of the operator, $P_\ell(t)$ may again fall back to less than $P_\ell^{max}$ before reaching $P_\ell^{out}$ and the protection device returns to the idle state. By following the assumptions made by Zima and Andersson \cite{Zima:2005} the probability for the line outage increases linearly with the power flow,
being zero below $P_\ell^{max}$. Thus, we assume $P_\ell^{out}$ being uniformly distributed in the interval $\left[ P_\ell^{max}, \beta P_\ell^{max} \right]$ with $\beta$=1.4. The line is either reconnected if the phase angle difference $\theta_\ell(t)=\theta_a(t)-\theta_b(t)$ becomes smaller than the preset value
$\theta_\ell^{max}= \eta x_{ab} P_\ell^{max}$ or after a time delay of $\Delta t_{\ell}^e$ which models the time until a manual attempt to re-close the breakers. The parameter $\eta$ is used to avoid an immediate recurrence of the overload, potentially resulting in a persistently repeating state change cycle, and is set to $\eta$=0.9. Analogous to the probabilistic failure model of the generating units (equations (\ref{eq:genF}-\ref{eq:genp})), the time to
permanent outage and the time to repair follow an exponential distribution with failure rate $\lambda_\ell$ and $\mu_\ell$ respectively.

\subsubsection{Busbars}
%\vspace{-0.2cm}
Every busbar object continuously calculates its phase angle $\theta_{a}(t)$ relative to its neighboring busbars:

\begin{equation}\label{eq:DCflow}
\theta_a(t)= \frac{P_a^{tot}(t)+ \displaystyle\sum_{b \in \Omega_a}\big(x_{ab}^{-1} \theta_b(t) \big)}{\displaystyle\sum_{b \in
\Omega_{a}}x_{ab}^{-1}}
\end{equation}
where $P_a^{tot}(t)= \sum_{j\in a}{P_j(t)}-\sum_{i\in a}{D_i(t)}$ is the net power injection at busbar $a$, to which several loads and
generating units might be connected. This distributed approach allows avoiding time consuming matrix calculations in case of network decompositions
and restorations due to line outages and reconnections. Potential random outages of busbars are not considered.

\subsubsection{Grid operator}

The grid operator becomes active in case of transmission line overload contingencies, trying to remove the overload by redispatching the generators or
by shedding load if necessary. The basic model for the operator behavior is illustrated for the overload of a tie-line between two control areas, see figure \ref{fig:gridOperator}. If a tie-line becomes overloaded the protection device sends an alarm message to the two operators of both control areas (compare figure \ref{fig:line}).
Having the alarm received the neighboring operators try to contact each other with a time delay $\Delta t_d^c$. The operator which has been assigned
responsible for the line then needs some time to find a solution to the overload problem, which is modeled by a time delay $\Delta t_d^r$. 

\begin{figurehere}
\centering
\includegraphics[width=3.5in]{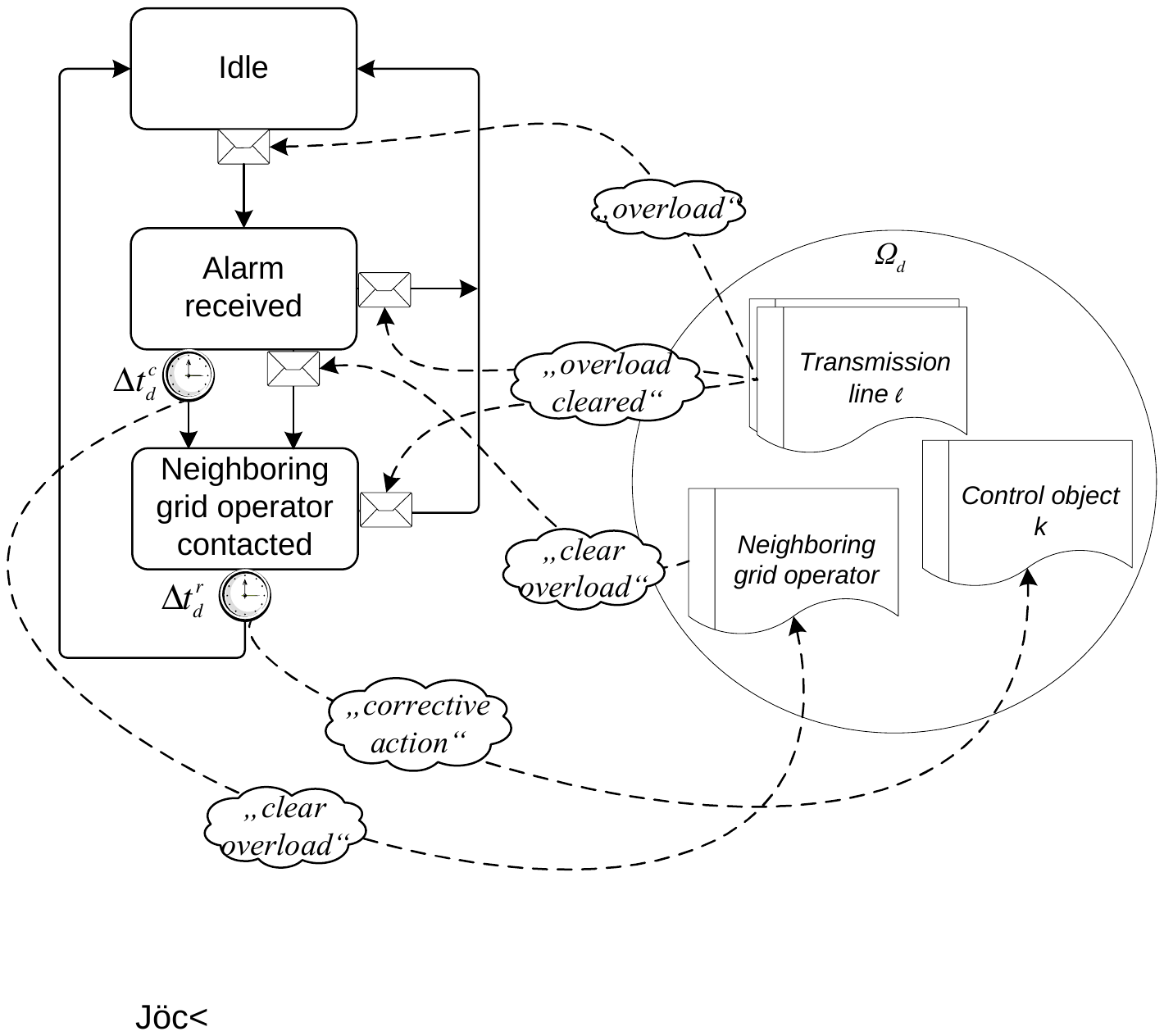}
\caption{Finite state machine for the grid operator} \label{fig:gridOperator}
\end{figurehere}
\vspace{+0.2cm}

The corrective action
to remove the overload is subsequently formulated as a conventional optimal power flow (OPF) problem \cite{Wood:1996} and implemented within the control object by using the linear programming (LP) method minimizing potential load shedding, $\Delta D_i$, and the change in generation, $\Delta P_j$, subject to the
transmission line constraints and the power balance:
%\vspace{-0.3cm}

\begin{equation}\label{eq:OPF}
\min z = \displaystyle\sum_{a=1}^{n_B}{\bigg( \omega_a \Big(\displaystyle\sum_{j \in a}{|\Delta P_j}| + W \displaystyle\sum_{i \in a} {\Delta D_i}
\Big) \bigg)}
\end{equation}
%\vspace{-0.2cm} 
subject to
\begin{equation}\label{eq:balance}
\sum_{j=1}^{n_G} {\Delta P_j} = -\sum_{i=1}^{n_L} {\Delta D_i}
\end{equation}
%\vspace{-0.2cm}
\begin{equation}\label{eq:loadReconnection}
-P_j(t)\leq \Delta P_j \leq P_j^{max} - P_j(t)
\end{equation}
%\vspace{-0.2cm}
\begin{equation}\label{eq:loadReconnection}
0 \leq \Delta D_i \leq D_i(t)
\end{equation}
%\vspace{-0.2cm}
\begin{equation}\label{eq:lineConstraints}
\bigg | P_\ell(t)+\displaystyle\sum_{a=1}^{n_B}{\bigg( a_a^\ell(t)} \Big(\sum_{j \in a} {\Delta P_j} + \sum_{i \in a} {\Delta D_i} \Big) \bigg) \bigg
|\leq \xi P_\ell^{max}
\end{equation}
%\vspace{-0.2cm}
%
where $\omega_a$ is the busbar specific distance weighting factor and set to $\omega_a$=1 for the two busbars at each end of the overloaded line,
$\omega_a$=10 for the busbars being one line further away and $\omega_a$=100 for all other busbars within the overall system. The weighting factor
$W$=10000 lets partial load shedding be more expensive relative to the generator redispatch. The linear line sensitivity factor
$a_a^\ell(t)=\frac{dP_\ell}{dP_a^{tot}}$ with respect to busbar $a$ is dependent on the network connectivity at the model time $t$ and is calculated
using the conventional matrix method as described in \cite{Wood:1996}. Equation (\ref{eq:lineConstraints}) holds for all lines within the two neighboring control areas. Similar to the model for the reconnection of a failed line, the parameter $\xi$ is used to delay the potential recurrence of an overload and set to $\xi$=0.8.  

The procedure for line overloads within a single control area is basically the same, but without the interaction of the operators and by restricting the load control variables $\Delta D_i$ to the busbars and equation (\ref{eq:lineConstraints}) to the transmission lines within the control area.
In order to prioritize the generator redispatch within the control area concerned, the distance weighting factor is set to $\omega_a$=1 for busbars inside and $\omega_a$=100 for busbars outside the control area.

\subsection{System Splitting}\label{section:Splitting}

The splitting of the network due to transmission line outages usually leads to an imbalance between load and generation within the separated
subsystems. Further, depending on the total inertia within the separated parts, on the frequency control performance and the protection device
behavior of the generators, and on implemented automatic load shedding procedures, this imbalance comes along with a frequency deviation potentially
leading to stability problems \cite{Anderson:1992}. The consequences range from small load losses to a total collapse of the subsystem (e.g.,
\cite{USBO:2004}). In order to include load outages as a consequence of a network splitting while avoiding a complicated model with a high amount of
parameters to be estimated we make use of a highly simplified approach. Thereby, an excess of demand within a separated subsystem leads to the
immediate disconnection of a minimum number of randomly selected loads so that the excess is at least reduced to zero. An excess of generation
leads to the immediate reduction of the generator outputs in order to reestablish the balance and implies no load outages. This strong simplification might be inadequate to represent the real system behavior and the amount of disconnected load thus has to be viewed as a rather indicative value for
the system vulnerability regarding the splitting of the network.

\subsection{Blackout Frequency Distributions}

By means of long-term simulations (i.e. repeatedly over the operation period of one year) the necessary statistical data is gathered for the
calculation of conventional reliability indices such as the Expected Energy Not Supplied (EENS). Moreover, frequency distributions of expected
blackouts per year are derived. Therefore, let $X$ be a random variable counting the number of blackouts per year greater than a specified size $C$.
The size is thereby classified by the unserved energy or the maximum amount of demand not being supplied in the course of an event. The expectation
$E(X)$ is approximated by generating $N$ realizations of $X$ and calculating their empirical mean, which represents the observed complementary
cumulative frequency of events related to one year, denoted by $\hat{F}_c(C)$:

\begin{equation}\label{eq:CCBF}
E(X)\approx \frac{1}{N}\displaystyle\sum_{i=1}^N X_i \equiv \hat{F} _c(C)
\end{equation}
where $N$ denotes the number of simulated years.

Assuming that $X$ follows a Poisson distribution, the confidence interval for $E(X)$ can be constructed by using the central Chi-square distribution \cite{Cowan:1998}:

\begin{equation}\label{eq:CI}
\gamma = 1- \alpha = Pr \Bigg[ \frac{1}{2N}\chi^2_{f^*;\alpha/2}\leq E(X) \leq \frac{1}{2N}\chi^2_{f;1-\alpha/2} \Bigg]
\end{equation}
where $\gamma$ is the confidence level, $\alpha$ is the probability of error, and $f^* = 2 \sum_{i=1}^N{X_i}$ and $f=2(\sum_{i=1}^N{X_i}+1)$ are the
degrees of freedom. The blackout events can further be classified into the three outage causes as implemented in the model:%\vspace{-0.2cm}
\begin{itemize}
\item Generation inadequacy: \newline Loads are disconnected as not enough generation capacity is available to cover the actual demand within the
overall system or within a previously separated subsystem. \item System splitting: \newline Loads are disconnected as a consequence of the
separation of the system. \item Operator intervention: \newline Load is partially shed in order to remove transmission line overloads.
\end{itemize}

\section{Case Studies}

\subsection{Application to the IEEE Reliability Test System 1996}

\subsubsection{System layout and model parameters}
%\vspace{-0.3cm}

The three-area IEEE Reliability Test System 1996 (RTS-96) has 73 busbars, 120 transmission lines and 96 generating units \cite{IEEERTS:1996}. We use
the year-long load data with an hourly resolution provided in \cite{IEEERTS:1996} for modeling the demand trajectories $D_i(t)$. The three areas have
a base case peak load $D_{k,0}^{max}$ of 2850 MW each and are treated as three single control areas with three corresponding control objects and grid operators. The
priorities given to the different generator types are shown in table \ref{tab:priorities}.

\vspace{0.2cm}
\begin{tablehere} \centering
\begin{tabular}{| c | c | c |}
\hline \hline
Unit Type & $P_j^{max}$ [MW] & Priority\\
\hline
Hydro & 50 & 1\\
Nuclear & 400 & 2\\
Coal/Steam & 350 & 3\\
Coal/Steam & 155 & 4\\
Coal/Steam & 76 & 5\\
Oil/Steam & 197 & 6\\
Oil/Steam & 100 & 7\\
Oil/Steam & 12 & 8\\
Oil/CT & 20 & 9\\

\hline \hline
\end{tabular}
\vspace{0.2cm}
\caption{Dispatch priorities for the generating units} \label{tab:priorities}
\end{tablehere}
\vspace{+0.2cm}

The failure and repair rates for the generators and the transmission lines are taken from \cite{IEEERTS:1996}. The parameter value for the time until the manual attempt to re-close the breaker of a disconnected line is assumed to be $\Delta t_{\ell}^e$=1h. Regarding the operator model $\Delta t_d^c$ is set to 2min.
%\vspace{-0.1cm}

\subsubsection{Computational results}
%\vspace{-0.3cm}

The results of two parameter variation studies are presented and discussed: 1) the sensitivity of the blackout frequency to an increase of the system
loading without any operator intervention and 2) the influence of the operator response time on the Expected Energy Not Supplied (EENS). Concerning the
first experiment we increment the system loading level $L=D_{k}^{max}/D_{k,0}^{max}$. The maximum generator outputs, $P_j^{max}$, are augmented by the same factor. In order to gain statistically significant results (i.e. $N$$\approx$1000) about 50 hours of simulation time are needed on a single desktop computer (Dell Optiplex GX260 with a Pentium 4 CPU of 2.66GHz and 512MB of RAM). This time was considerably reduced by running several simulations in parallel. Figure \ref{fig:CCBF} shows the resulting complementary cumulative blackout frequencies with respect to the unserved energy per event, $\hat{F}_c(C_E)$, for four different values of
$L$. Regarding the two lower system loading levels ($L$=1.0 and $L$=1.1) the observed complementary cumulative frequencies follow approximately an
exponential curve. However, increasing $L$ to 1.2 already leads to a remarkable increase of large events, while the shape of the curve in the range
of the smaller events (up to about $10^3$ MWh) stays qualitatively the same. The value of $L$=1.37 represents the maximum system loading level where no
line overloads would occur without any stochastic component outages. This loading level can be characterized by a high frequency of large blackouts
predominantly in the range between $10^4$ MWh and $10^5$ MWh.

\begin{figurehere}
\centering
\includegraphics[width=3.2in]{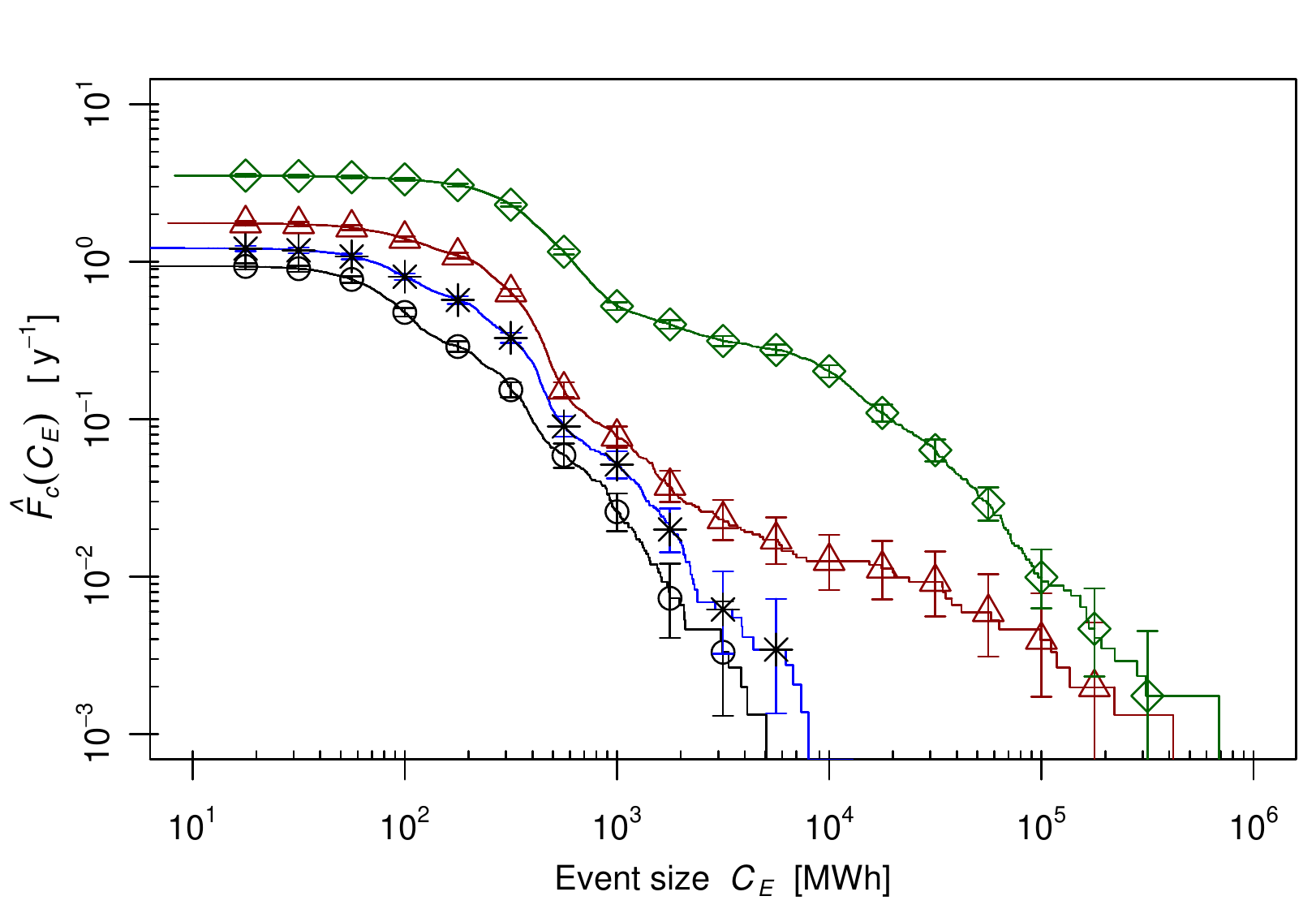}
\caption{Complementary cumulative blackout frequencies for four different system loading levels $L$=1.0, 1.1, 1.2 and 1.37 (circles, stars, triangles
and diamonds, respectively) without operator intervention. The error bars indicate the 90\% confidence interval.} \label{fig:CCBF}
\end{figurehere}
\vspace{0.2cm}

In order to further analyze the differences between the overall frequency
curves the distributions of the underlying power outage causes have to be considered. The logarithmic histograms of figure \ref{fig:hist} report the
impact of increasing the system loading from $L$=1.0 to $L$=1.37 on the absolute frequency of blackouts $f(C_E)$ caused by generation inadequacy
(left hand side) and system splitting (right hand side). System splitting is the predominant cause of the observed blackouts for both loading levels. In comparison to generation inadequacy the absolute frequencies for this outage mode show a stronger increase and a stronger shift towards larger events when it comes to an increase of
the system loading. Hence, the substantial increase of large blackouts as shown in figure \ref{fig:CCBF} is mainly the result of an increased frequency of line overloads and subsequent system splitting.

\begin{figurehere}
\centering
\includegraphics[width=3.2in]{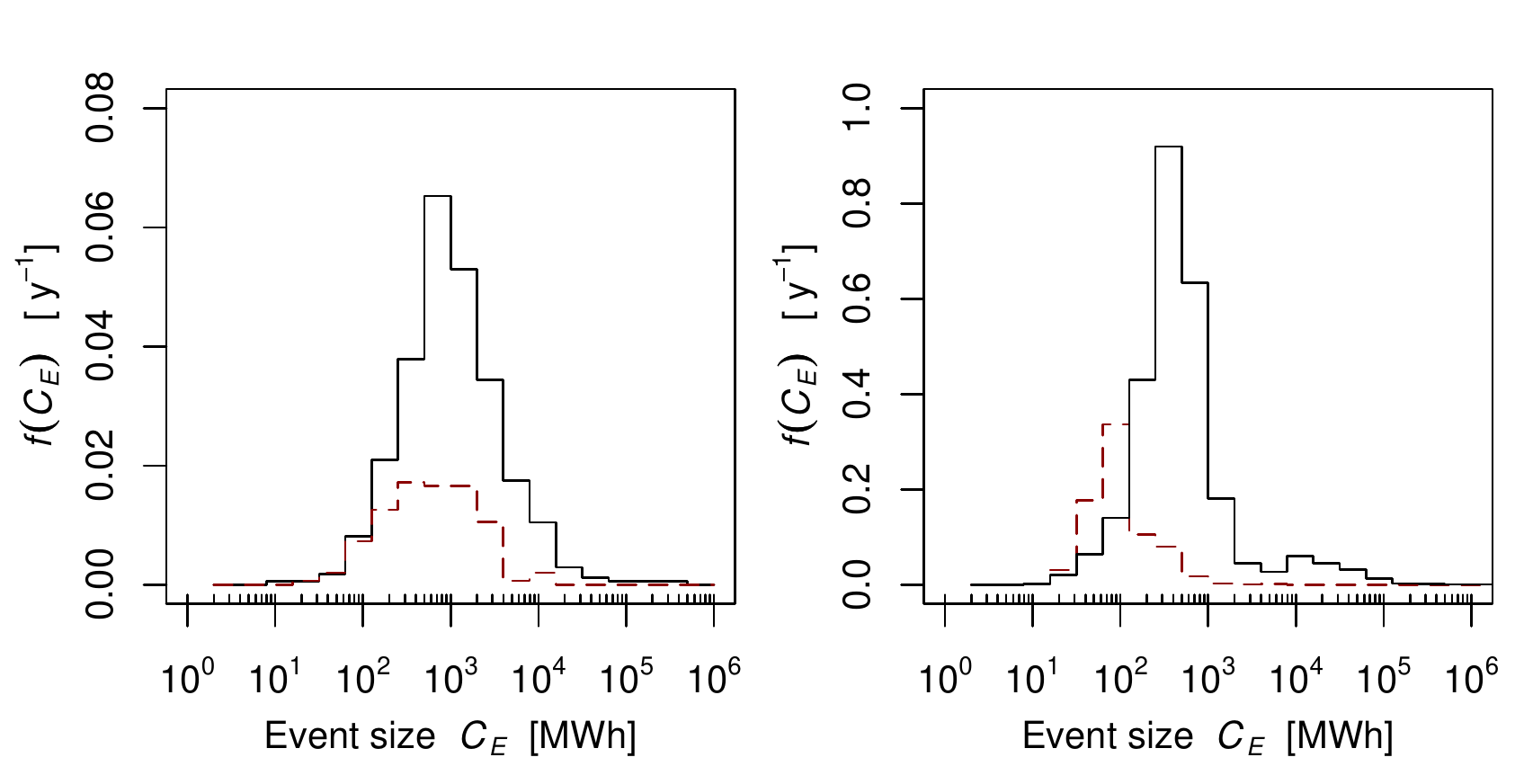}
\caption{Impact of increasing the system loading from $L$=1.0 (dashed line) to $L$=1.37 (continuous line) on the absolute frequencies of blackouts
caused by generation inadequacy (left) and system splitting (right).} \label{fig:hist}
\end{figurehere}
\vspace{0.2cm}

The results of our second parameter variation study are presented in figure \ref{fig:operator}, showing the influence of the
operator response time $\Delta t_d^r$ on the EENS broken down into the different outage causes for the system loading level $L$=1.37. For the interpretation of the results it should be reminded that thermal aspects of the line overloads are not taken into consideration. Under our model assumptions an operator intervention with a delay of 5 hours after the occurence of the overload still reduces the EENS due to system splitting by about 30\%. On the other hand, an increase of the response time from 15min to 30min leads to a significant increase of the EENS due to system splitting of about 26\%. The EENS due to generation inadequacy is increasing with the response time as the system is more often separated which, in turn, reduces the redundancy of the generators within the splitted subsystems. The values for the EENS due to the operator intervention are negligible.

\begin{figurehere}
\centering
\includegraphics[width=3.1in]{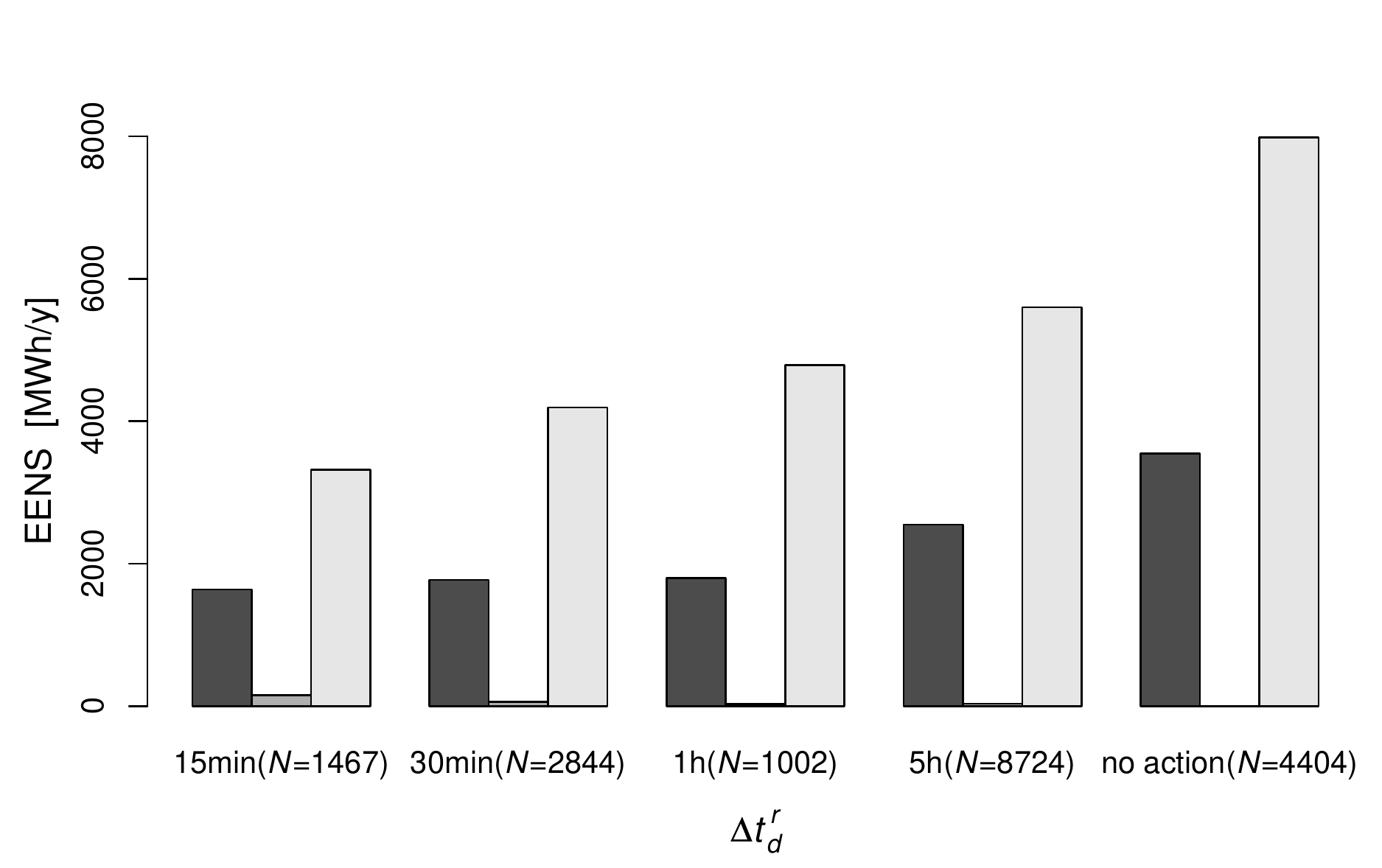}
\caption{Influence of the operator response time on the EENS due to generation inadequacy (left, black bar), operator action (middle, dark-grey bar) and system splitting (right, light-grey bar) for $L$=1.37. }
\label{fig:operator}
\end{figurehere}
\vspace{0.2cm}

\subsection{Application to the Swiss High-Voltage Grid}
\vspace{-0.0cm}

\subsubsection{System layout and model parameters}
%\vspace{-0.2cm}

The Swiss electric power system consists of a single control area with an annual energy consumption of $62.1\cdot10^3$ GWh and a peak load of about
10 GW. The energy production and installed capacity total to $59.4\cdot10^3$ GWh and 12 GW respectively, consisting of 42.2\% nuclear, 52.4\% hydro
and 5.4\% conventional thermal generation \cite{BFE:2006}. The number of components as used in our model for the 380/220 kV transmission grid are
shown in \mbox{table \ref{tab:SwissNum}.}

\begin{tablehere} \centering
\begin{tabular}{| c | c | c | c | c |}
\hline \hline
$n_L$ & $n_G$ & $n_T$ & $n_B$ & $n_K$\\
\hline
99 & 34 & 229 & 161 & 1\\

\hline \hline
\end{tabular}
\vspace{0.2cm}
\caption{Number of components of the Swiss system} \label{tab:SwissNum}
\end{tablehere}
\vspace{+0.2cm}

Based on a particular system snapshot taken on a January morning the fluctuating power injections $P_a^{tot}(t)$ at the different nodes are derived
by using publicly available statistical data \cite{BFE:2006}. For each hydro power generator a different production capacity is assigned for the
winter half-year and the summer half-year respectively. The failure and repair rates for all hydro generators are equally set to
$\lambda_j=4.42y^{-1}$ and $\mu_j=0.05h^{-1}$, and for all nuclear units to $\lambda_j=3y^{-1}$ and $\mu_j=0.027h^{-1}$. Regarding the transmission
lines the failure model parameters are chosen as $\lambda_\ell= 0.234 y^{-1}$ and $\mu_\ell=0.056h^{-1}$. As the phase shifting transformers have
considerable influence on the power flows corrective injections were calculated for the nodes adjacent to a phase shifting transformer. The energy
exchange with the neighboring countries is modeled by independent positive or negative power injections at the surrounding boundary nodes. The parameter values for the time until the manual attempt to re-close the breaker of a disconnected line and for the operator response time are assumed to be $\Delta t_{\ell}^e$=1h and $\Delta t_d^r$=15min, respectively.

\subsubsection{Computational results}

It should be noted that the intention of the analysis was primarily to investigate the applicability of the proposed modeling method to a real
system. The computational results thus make no claim to quantify the reliability of the Swiss high-voltage grid in absolute terms.

The estimated
blackout frequencies and the histogram of the different outage causes both with respect to the unserved energy per event are depicted in figure
\ref{fig:SwissA}. The model potentially overestimates the duration of the events and thus the unserved energy as switching operations on lower
voltage levels for the reconnection of deenergized loads and the possibility to import extra power from neighboring countries to overcome generation
shortages are not taken into consideration.

\begin{figurehere}
\centering
\includegraphics[width=3in]{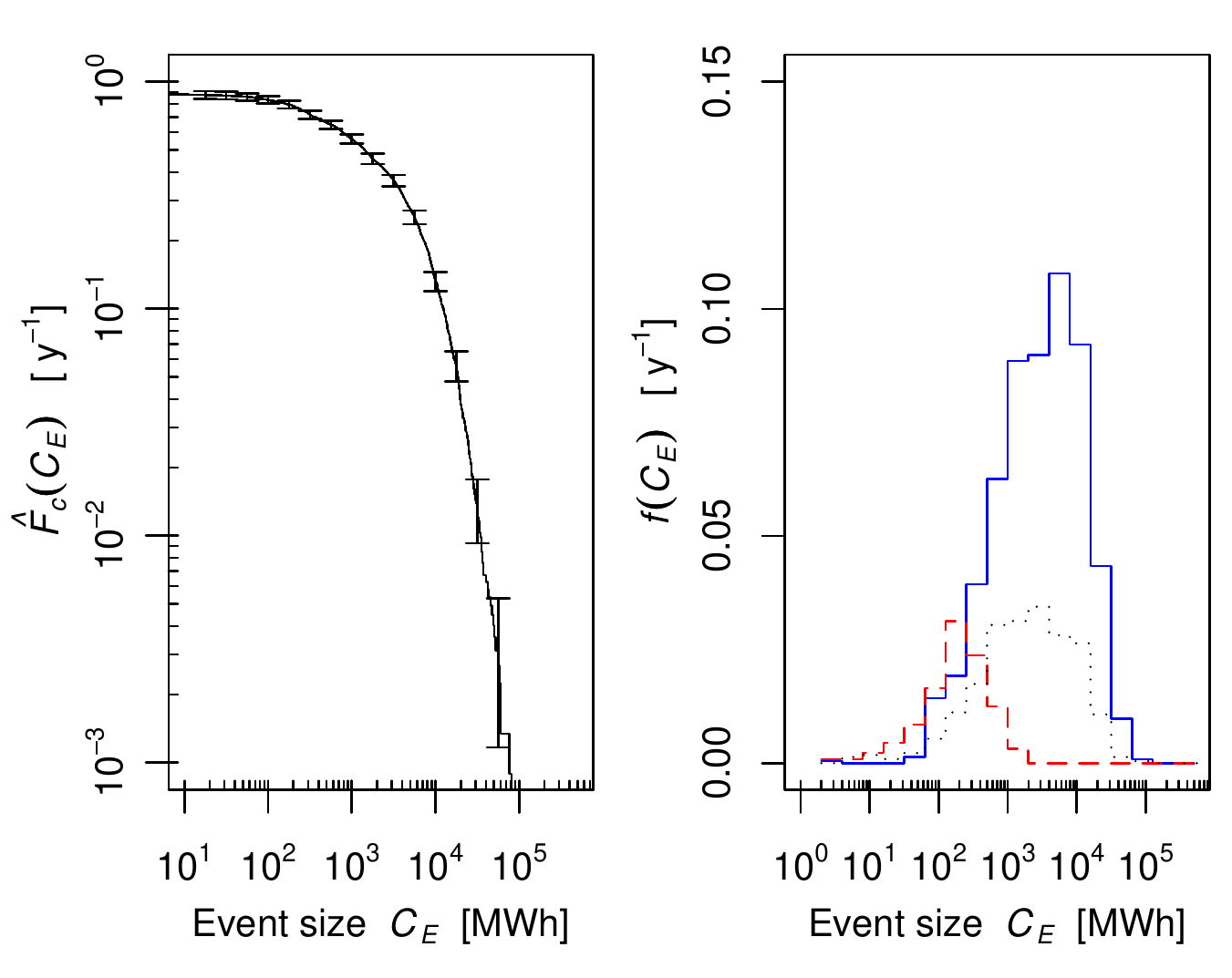}
\caption{Left: estimated blackout frequencies for the Swiss system with respect to the unserved energy. The error bars indicate the 90\% confidence
interval. Right: histogram indicating the distribution of the outages due to generation inadequacy (continuous line), system splitting (dotted line)
and load shedding for line overload removal (dashed line).} \label{fig:SwissA}
\end{figurehere}

The complementary cumulative blackout frequency follows an exponential curve. Generation inadequacy is the dominant factor regarding the larger events while load shedding for line overload relief becomes important in the range of the smaller events. The influence of load
disconnections due to system splitting is significant but the frequency of this outage cause never exceeds the frequency of load disconnections due to
generation inadequacy or load shedding due to the operator action. Hence, under our model assumptions, it can be concluded that the system reliability is
somewhat more sensitive to generation outages than to transmission line failures. 

The benefit of the operator response to line overloads is
shown in figure \ref{fig:SwissB} where the frequencies of the events with and without operator action are compared. The event size is thereby measured by the maximum unserved demand.
%\vspace{-0.5cm}

\vspace{+0.1cm}
\begin{figurehere}
\centering
\includegraphics[width=3.3in]{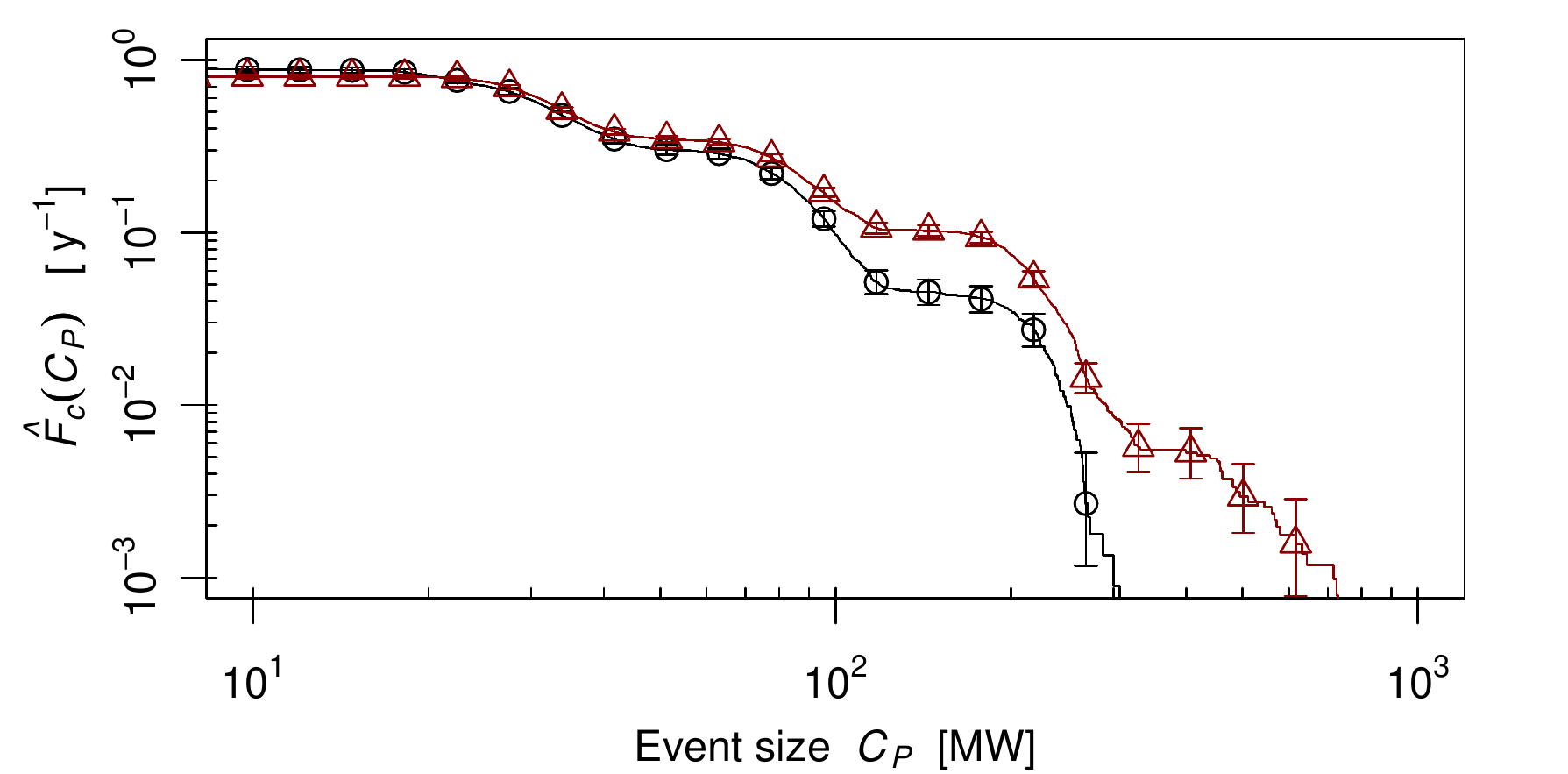}
\caption{Blackout prevention due to operator response to line overloads. Triangles: no operator intervention, circles: operator intervention with $\Delta t_d^r = 15$min. The error bars indicate the 90\% confidence interval.} \label{fig:SwissB}
\end{figurehere}
\vspace{+0.1cm}

The impact of the operator intervention becomes significant in the range of the larger events where a high fraction of blackouts with a size greater
than 200 MW is prevented. A large number of disconnected loads due to system splitting and thus a high value for the unserved demand generally needs
a high number of subsequently disconnected lines due to overload. Such a sequence of events potentially gives the operator a higher chance to
intervene in comparison to a disconnection of a single load due to the outage of a few lines without further cascading failures.

The relative overload frequencies for each transmission line, $h_\ell$, are reported in figure \ref{fig:Swisslines}. About 15\% of all overload contingencies are occurring on only two lines. Furthermore, several groups of adjacent lines can be identified as being
prone to overloads, helping to highlight the most critical system regions.

\begin{figurehere}
\centering
\includegraphics[width=3.2in]{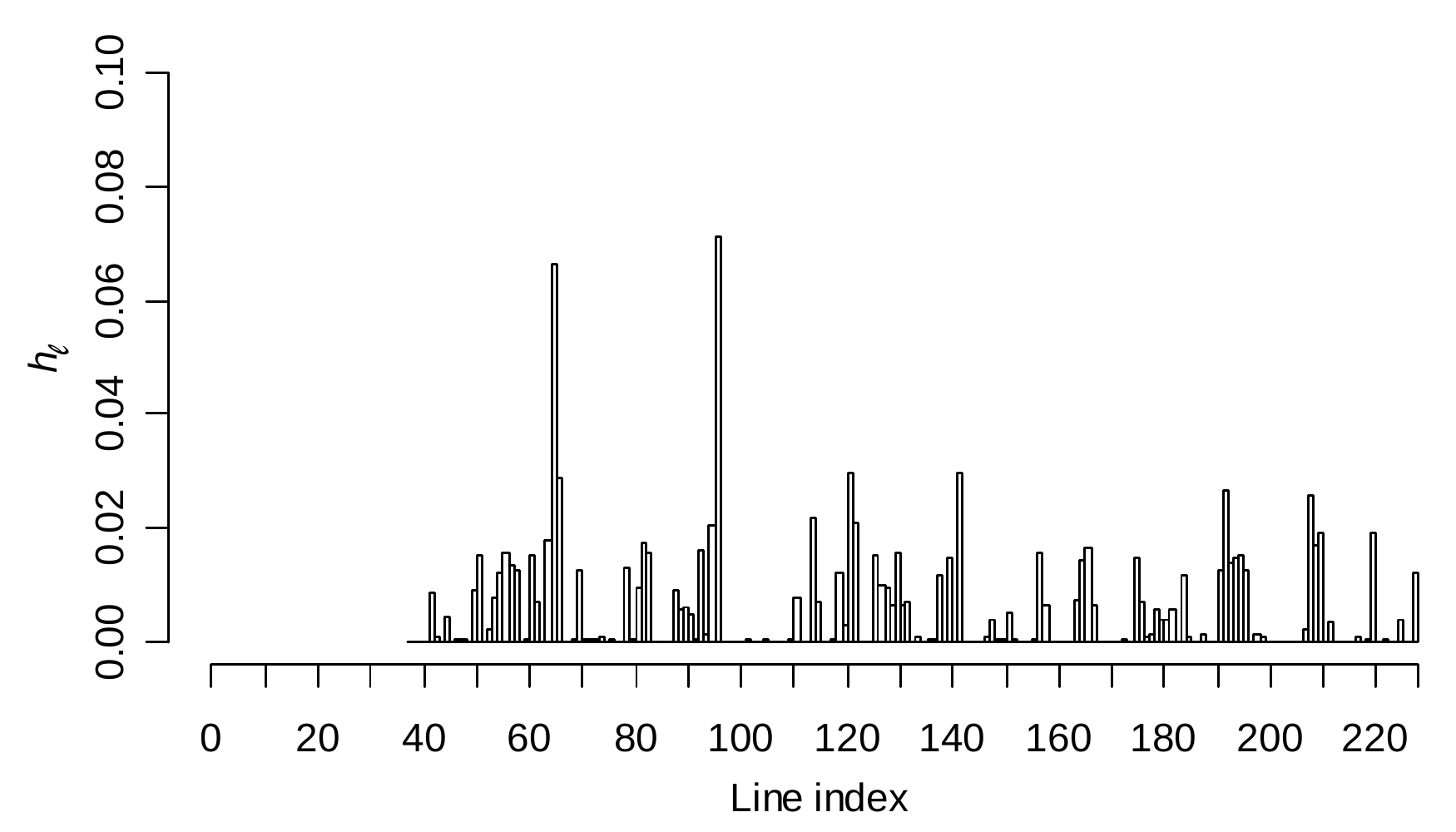}
\caption{Relative frequency of transmission line overloads ($N$=2242)} \label{fig:Swisslines}
\end{figurehere}
\vspace{+0.2cm}

\section{Conclusions}

We presented an object-oriented hybrid modeling framework for a comprehensive reliability analysis of electric power systems. The main advantages are the explicit integration of highly nonlinear, time-dependent effects and the possibility to include non-technical factors. The chosen level of modeling detail allows analyzing a multitude of different (time-dependent) reliability aspects such as the identification of weak points and the assessment of system upgrades. Although several model refinements need to be further developed, the results of the case studies performed on the IEEE RTS-96 and on a model of the Swiss high-voltage grid confirm the applicability of the approach with respect to mid-period power system planning purposes. Optimizing the technical implementation of the models \cite{Schlapfer:2011} together with the evolution of both hard- and software will fasten up the simulation speed. Gaining experience in applying the proposed approach will give insight in the parameters to be used, thus lessen the problem of the high number of parameters to be estimated.

\section*{ACKNOWLEDGEMENTS}

The authors would like to thank "swiss\emph{electric} research" for co-financing the work, Swissgrid AG for the fruitful collaboration and for
providing the operational data of the Swiss electric power system, and Walter Sattinger (Swissgrid AG) for his helpful feedback on the manuscript.

\end{multicols}
\end{document}